\documentstyle[12pt]{article}
\begin{document}
\begin{titlepage}
\begin{flushright}
IC/2001/81\\
hep-th/0107174
\end{flushright}
\vspace{10 mm}

\begin{center}
{\Large Brane World in a Topological Black Holes \\
in Asymptotically Flat Spacetime}

\vspace{5mm}

\end{center}
\vspace{5 mm}

\begin{center}
{\large Donam Youm\footnote{E-mail: youmd@ictp.trieste.it}}

\vspace{3mm}

ICTP, Strada Costiera 11, 34014 Trieste, Italy

\end{center}

\vspace{1cm}

\begin{center}
{\large Abstract}
\end{center}

\noindent

We study static brane configurations in the bulk background of the topological 
black holes in asymptotically flat spacetime.  We find that such  
configurations are possible even for flat black hole horizon, unlike the AdS 
black hole case.  We construct the brane world model with an orbifold 
structure $S^1/{\bf Z}_2$ in such bulk background.  We also study massless 
bulk scalar field.

\vspace{1cm}
\begin{flushleft}
July, 2001
\end{flushleft}
\end{titlepage}
\newpage
\begin{sloppypar}

Recently, there has been active interest in the possibility that our 
four-dimensional universe may be a three-brane embedded in higher-dimensional 
spacetime \cite{add,aadd,rs1,rs2}, as such brane world scenarios might 
provide possible solutions to the hierarchy and the cosmological constant 
problems.  In the scenario proposed by Randall and Sundrum (RS) 
\cite{rs1,rs2}, the extra spatial dimension needs not be compact for 
reproducing four-dimensional gravity on the brane due to the strongly warped 
bulk spacetime and the fine-tuned brane tension is required for solving the 
hierarchy and the cosmological constant problems.

In particular, the brane universes in the bulk of the five-dimensional 
AdS black holes \cite{kra,ida,ceg} and higher-dimensional branes \cite{kk} 
in string theories have been studied.  In such models, the motion of the 
probe brane in the bulk of the source brane is responsible for the expansion 
of the brane universe on the probe brane.  When the equation of state for 
the matter field on the brane takes a special form, the static brane 
configuration, in which the probe brane remains static in the bulk, is 
possible.  Such static brane configurations may be regarded as 
generalizations of the static RS model \cite{rs1,rs2} to different bulk 
spacetimes.  In Refs. \cite{gjs,kmp}, it was found out that such static 
brane configuration is possible even without matter fields on the probe 
brane for some bulk spacetimes.  Such result was generalized to the case 
of the topological AdS black holes in Ref. \cite{br}.  

In this paper, we study the static brane configurations in the bulk of the 
topological black holes in asymptotically flat spacetime.  Contrary to the 
previous belief that black holes in asymptotically flat spacetime should 
have spherical horizon \cite{haw,fsw}, new class of dilatonic $p$-brane 
solutions in asymptotically flat spacetime where horizon is the maximally 
symmetric space with any values of the curvature parameter was constructed 
\cite{gqtz}.  In particular, uncharged 0-brane solution with flat horizon 
can be transformed to the self-tuning flat domain wall solution 
\cite{adks,kss} when the dilaton coupling parameter takes a special value.  
We find that the static brane configuration is possible for any horizon 
geometries, unlike the case of the topological AdS black holes studied in 
Ref. \cite{br}.  We obtain the fine-tuned value of the brane tension and 
the stabilized location of the probe brane in terms of the parameters of the 
bulk black hole solution.  We also study massless scalar field in such bulk 
background.

We begin by discussing the new class of dilatonic $p$-brane solutions 
constructed in Ref. \cite{gqtz}.   For the purpose of putting the solutions 
into more convenient forms, we choose to parameterize the action for the 
$p$-brane solutions in the following form:
\begin{equation}
S_p={1\over{2\kappa^2_D}}\int d^Dx\sqrt{-G}\left[{\cal R}-{4\over{D-2}}
(\partial\phi)^2-{1\over{2\cdot(p+2)!}}e^{2a_p\phi}F^2_{p+2}\right],
\label{pbrnact}
\end{equation}
where $\kappa^2_D$ is the $D$-dimensional gravitational constant, $\phi$ 
is the dilaton field, $F_{p+2}$ is the field strength of a $(p+1)$-form 
potential $A_{p+1}$ and $a_p$ is the dilaton coupling parameter.  The most 
general form of the action with arbitrary coefficients in front of the 
kinetic terms, considered in Ref. \cite{gqtz}, is related to this action 
just through rescaling of $\phi$ and $A_{p+1}$ with constant factors.  In 
terms of such new parametrization, the solution for the topological 
non-extreme dilatonic $p$-brane with the longitudinal coordinates ${\bf x}
=(x^1,\dots,x^p)$, constructed in Ref. \cite{gqtz}, takes the following form:
\begin{eqnarray}
G_{MN}dx^Mdx^N&=&-h^{{{4(n-1)}\over{(p+n)\Delta_p}}-1}_-h_+dt^2+
h^{{4(n-1)}\over{(p+n)\Delta_p}}_-d{\bf x}\cdot d{\bf x}+
h^{-{{4(p+1)}\over{(p+n)\Delta_p}}-{{n-3}\over{n-1}}}_-h^{-1}_+dr^2
\cr
& &\ \ \ \ \ 
+h^{-{{4(p+1)}\over{(p+n)\Delta_p}}+{2\over{n-1}}}_-r^2ds^2_{n,k},
\cr
e^{2\phi}&=&h^{-{{(p+n)a_p}\over\Delta_p}}_-,\ \ \ \ \ \ \ \ \ \ 
A_{tx^1\dots x^p}={2\over\sqrt{\Delta_p}}{{m\sinh\alpha\cosh\alpha}
\over r^{n-1}},
\cr
h_+&=&s(r)\left(1-{{m\cosh^2\alpha}\over r^{n-1}}\right),\ \ \ \ \ \ \ 
h_-=\left|k-{{m\sinh^2\alpha}\over r^{n-1}}\right|,
\cr
\Delta_p&=&{{(p+n)a^2_p}\over 2}+{{2(p+1)(n-1)}\over{p+n}},
\label{toppbrn}
\end{eqnarray}
where $n\equiv D-p-2$, $s(r)\equiv{\rm sgn}(k-m\sinh^2\alpha/r^{n-1})$ 
and $ds^2_{n,k}$ is the metric for the maximally symmetric $n$-dimensional 
space with constant curvature, which we parametrize as
\begin{equation}
ds^2_{n,k}=\left(1+\textstyle{k\over 4}\delta_{ij}y^iy^j\right)^{-2}
[(dy^1)^2+\dots+(dy^n)^2],
\label{maxsymmet}
\end{equation}
and $k=0,\pm 1$ is the curvature parameter.  With our parametrization it 
is manifest that the curvature singularity at $r=r_-=(m\sinh^2
\alpha)^{1/(n-1)}$ for the $k=1$ case is always covered by a horizon at 
$r=r_+=(m\cosh^2\alpha)^{1/(n-1)}$, provided $m>0$.  In this paper, we 
consider the $p=0$ case, for which the solution takes the form:
\begin{eqnarray}
G_{MN}dx^Mdx^N&=&-h^{{{4(n-1)}\over{n\Delta}}-1}_-h_+dt^2+
h^{-{4\over{n\Delta}}-{{n-3}\over{n-1}}}_-h^{-1}_+dr^2+
h^{-{4\over{n\Delta}}+{2\over{n-1}}}_-r^2ds^2_{n,k},
\cr
e^{2\phi}&=&h^{-{{na}\over\Delta}}_-,\ \ \ \ \ \ \ \ \ \ 
A_t={2\over\sqrt{\Delta}}{{m\sinh\alpha\cosh\alpha}\over r^{n-1}},
\cr
h_+&=&s(r)\left(1-{{m\cosh^2\alpha}\over r^{n-1}}\right),\ \ \ \ \ \ \ 
h_-=\left|k-{{m\sinh^2\alpha}\over r^{n-1}}\right|,
\cr
\Delta&=&{{na^2}\over 2}+{{2(n-1)}\over n},
\label{topbh}
\end{eqnarray}
where now $D=n+2$ and $a:=a_0$ is the dilaton coupling parameter for the 
$U(1)$ gauge field $A_M$.  We may regard this topological 0-brane as 
describing bulk spacetime of a brane world with the worldvolume 
coordinates $(t,y^1,...,y^n)$ and the transverse coordinate $r$.  However, 
unlike the RS model \cite{rs1,rs2}, such brane world model generally 
has asymmetrically warped spacetime, meaning that the space and the time 
coordinates have different warp factors, and as a result the Lorentz 
invariance is violated \cite{ceg}.  However, for the uncharged 0-brane 
solution with $k=0$, the bulk metric can become $(D-1)$-dimensional 
Poincar\'e invariant.  The uncharged solution in Ref. \cite{gqtz} takes the 
following form in our parametrization:
\begin{eqnarray}
G_{MN}dx^Mdx^N&=&-h^{{{4(n-1)}\over{n\Delta}}-1}_-h_+dt^2+
h^{-{4\over{n\Delta}}-{{n-3}\over{n-1}}}_-h^{-1}_+dr^2+
h^{-{4\over{n\Delta}}+{2\over{n-1}}}_-r^2ds^2_{n,k},
\cr
e^{2\phi}&=&h^{-{{na}\over\Delta}}_-,\ \ \ \ \ 
h_+={\rm sgn}\,h,\ \ \ \ \ h_-=|h|,\ \ \ \ \ h=k-{l\over r^{n-1}},
\label{uchrgdsol}
\end{eqnarray}
where $l$ is an integration constant.  Note, this uncharged $p$-brane 
solution cannot be obtained by taking the zero charge limit of the charged 
$p$-brane solution (\ref{toppbrn})
\footnote{This fact does not imply that our convenient parametrization 
(\ref{toppbrn}) for the charged topological $p$-brane solution is wrong.  
This can be easily seen by the fact that the usual uncharged black $p$-brane 
solution (with $k=1$) cannot be coordinate-transformed to take the form 
(\ref{uchrgdsol}) with $k=1$.}.  
With a choice of $l=-1$ and $k=0$, the metric (\ref{uchrgdsol}) has the 
$(D-1)$-dimensional Poincar\'e invariance, when the temporal and the spatial 
warp factors have the same $r$-dependence:
\begin{equation}
r^{-(n-1)\left[{{4(n-1)}\over{n\Delta}}-1\right]}=r^2\cdot r^{-(n-1)
\left[-{4\over{n\Delta}}+{2\over{n-1}}\right]}\ \ \ \ \ \ \ 
\Longrightarrow\ \ \ \ \ \ \ \Delta=4.
\label{poincrcndtn}
\end{equation}
For such case, the uncharged solution (\ref{uchrgdsol}) becomes the 
self-tuning domain wall solution constructed in Refs. \cite{adks,kss}, after 
the redefinition of the transverse coordinate $r$ \cite{gqtz}.  So, we 
see that the self-tuning domain wall solution can be embedded in string 
theories as an uncharged 0-brane solution with flat ($k=0$) horizon and 
$\Delta=4$.  Such uncharged 0-brane solution can be obtained by 
compactifying, for example, uncharged D6-brane, for which $\Delta_6=4$, 
on a compact Ricci flat manifold along its longitudinal directions.  

In this paper, we study the brane world on a probe $n$-brane in the bulk 
background of the topological dilatonic 0-brane (\ref{topbh}).   For such 
purpose, it is convenient to redefine the transverse coordinate in the 
following way \cite{gjs,kmp}:
\begin{equation}
dz=h^{-{2\over{n\Delta}}-{{n-3}\over{n-1}}}_-h^{-{1\over 2}}_+dr,
\label{redef}
\end{equation}
so that the bulk metric (\ref{topbh}) takes the form that resembles 
the RS domain wall metric except that the warp factor is asymmetric:
\begin{equation}
G_{MN}dx^Mdx^N=-A(z)^2dt^2+B(z)^2\sigma_{ij}(x)dx^idx^j+dz^2,
\label{newmet}
\end{equation}
where $A(z)$ and $B(z)$ are implicitly given by
\begin{equation}
A(z)=h^{{{2(n-1)}\over{n\Delta}}-1}_-h^{1\over 2}_+,\ \ \ \ \ \ \ 
B(z)=h^{-{2\over{n\Delta}}+{1\over{n-1}}}_-r,
\label{defab}
\end{equation}
with $r$ related to $z$ in the way defined in Eq. (\ref{redef}).  
We consider the following form of the $n$-brane action:
\begin{equation}
S_{\sigma}=-\int d^{n+1}x\sqrt{-g}f(\phi),
\label{brnact}
\end{equation}
where $g$ is the determinant of the induced metric $g_{\mu\nu}=\delta^M_{\mu}
\delta^N_{\nu}G_{MN}$ on the $n$-brane and $f(\phi)$ is an arbitrary function 
of $\phi$.  

We consider the static brane configuration without matter fields on 
the $n$-brane.  From the total action $S_{p=0}+S_{\sigma}$, we obtain 
the following Einstein's equations and equation of motion for the 
dilaton field:
\begin{eqnarray}
{\cal G}_{MN}&=&-\textstyle{2\over{D-2}}G_{MN}(\partial\phi)^2+
\textstyle{4\over{D-2}}\partial_M\phi\partial_N\phi+\textstyle{1\over 2}
e^{2a\phi}(F_{MP}F^{\ P}_N-\textstyle{1\over 4}G_{MN}F^2_2)
\cr
& &-\kappa^2_D\sqrt{g\over G}\delta^{\mu}_M\delta^{\nu}_Ng_{\mu\nu}f(\phi)
\delta(z-z_b),
\label{eqmtn1}
\end{eqnarray}
\begin{equation}
\textstyle{4\over{D-2}}\partial_M\left[\sqrt{-G}G^{MN}\partial_N\phi\right]
=\textstyle{a\over 4}e^{2a\phi}F^2_2+\kappa^2_D\sqrt{-g}f^{\prime}(\phi)
\delta(z-z_b),
\label{eqmtn2}
\end{equation}
where ${\cal G}_{MN}={\cal R}_{MN}-{1\over 2}G_{MN}{\cal R}$ is the Einstein 
tensor for the bulk metric $G_{MN}$, the constant $z_b$ is the location of 
the $n$-brane and $D=n+2$.  After the explicit expression for the bulk metric 
(\ref{newmet}) is substituted into these equations of motion, the $(z,z)$-, 
$(t,t)$- and $(i,j)$-components of the Einstein's equations and the dilaton 
equation of motion respectively take the following forms:
\begin{eqnarray}
-{{(D-2)(D-3)}\over 2}{k\over B^2}+(D-2){{A^{\prime}B^{\prime}}\over{AB}}
+{{(D-2)(D-3)}\over 2}{B^{\prime\,2}\over B^2}
\ \ \ \ \ \ \ \ \ \ \ \ \ \ \ 
\cr
={2\over{D-2}}(\partial_z\phi)^2-{3\over 8}{1\over A^2}e^{2a\phi}F^2_{tz},
\label{eqn1}
\end{eqnarray}
\begin{eqnarray}
{{(D-2)(D-3)}\over 2}{k\over B^2}-(D-2){B^{\prime\prime}\over B}
-{{(D-2)(D-3)}\over 2}{B^{\prime\,2}\over B^2}
\ \ \ \ \ \ \ \ \ \ \ \ \ \ \ 
\cr
={2\over{D-2}}(\partial_z\phi)^2+{3\over 8}{1\over A^2}e^{2a\phi}F^2_{tz}
+\kappa^2_Df(\phi)\delta(z-z_b),
\label{eqn2}
\end{eqnarray}
\begin{eqnarray}
{{D-3}\over 2}\left[-(D-4){k\over B^2}+(D-4){B^{\prime\,2}\over B^2}+
2{B^{\prime\prime}\over B}+2{{A^{\prime}B^{\prime}}\over{AB}}\right]+
{A^{\prime\prime}\over A}
\ \ \ \ \ \ \ \ \ \ \ \ \ \ \ 
\cr
=-{2\over{D-2}}(\partial_z\phi)^2-\kappa^2_Df(\phi)\delta(z-z_b),
\label{eqn3}
\end{eqnarray}
\begin{equation}
{4\over{D-2}}\partial_z(AB^n\partial_z\phi)=-{a\over 4}{1\over A^2}e^{2a\phi}
F^2_{tz}+\kappa^2_DAB^nf^{\prime}(\phi)\delta(z-z_b),
\label{eqn4}
\end{equation}
where the prime denotes the derivative w.r.t. the transverse 
coordinate $z$.  

We now construct the brane world model with an orbifold structure 
$S^1/{\bf Z}_2$ just like the RS1 model \cite{rs1}.  Following the 
procedure proposed in Ref. \cite{gjs}, we cut off the transverse space 
at the black hole horizon $z=z_H$ (corresponding to $r=r_+=(m\cosh^2
\alpha)^{1/(n-1)}$) and at the location $z=z_b$ of the $n$-brane.  The 
spacetime between these boundaries is referred to as the bulk.  We 
introduce a new transverse coordinate $\bar{z}\equiv z_b-z$ and identify 
the corresponding two points under the ${\bf Z}_2$-symmetry $\bar{z}\to
-\bar{z}$.  So, in this new coordinate the $n$-brane is located at 
$\bar{z}=0$ and the black hole horizon is located at orbifold points 
$\bar{z}=\pm|z_b-z_H|$, which are identified under the ${\bf Z}_2$-symmetry 
$\bar{z}\to-\bar{z}$.  We rewrite the bulk metric (\ref{newmet}) in terms 
of the new coordinates as
\begin{equation}
G_{MN}dx^Mdx^N=-A(|\bar{z}|)^2dt^2+B(|\bar{z}|)^2\sigma_{ij}(x)dx^idx^j
+d\bar{z}^2,
\label{blkmetnewcrd}
\end{equation}
where $A$ and $B$ are now functions of $|\bar{z}|$ due to the imposed 
${\bf Z}_2$-symmetry and $-|z_b-z_H|\leq\bar{z}\leq|z_b-z_H|$.  Although 
the metric components are continuous at $\bar{z}=0$, their first and 
second derivatives w.r.t. $\bar{z}$ are discontinuous and have 
$\delta$-function singularity at $\bar{z}=0$, respectively.  

By applying the boundary conditions on the first derivatives at $z=z_b$ 
resulting from the $\delta$-function terms in Eqs. 
(\ref{eqn2},\ref{eqn3},\ref{eqn4}), we can fix the parameters in the 
$n$-brane action (\ref{brnact}) and the (stabilized) location $r_b$ of 
the $n$-brane.  For the phenomenological relevance, from now on we consider 
the $k=0$ case, only
\footnote{The nontrivial static brane configuration is possible also when 
$k=\pm 1$.  The below equations (\ref{rel1}) and (\ref{drvrel}) continue to 
hold even for the $k=\pm 1$ cases.  However, the stabilized brane location 
$r_b$ take more complicated forms $r^{n-1}_b=m(2n-2+n\Delta-2\Delta)\cosh^2
\alpha\sinh^2\alpha/[(2n-2+n\Delta-2\Delta)\cosh^2\alpha-2n+2+\Delta]$ and 
$r^{n-1}_b=m(2n-2+n\Delta-2\Delta)\cosh^2\alpha\sinh^2\alpha/[(2n-2-n\Delta)
\cosh^2\alpha-2n+2+\Delta]$, respectively for $k=1$ and $k=-1$.}.  
By applying the boundary conditions resulting from Eqs. 
(\ref{eqn2},\ref{eqn4}) we obtain the following constraints on the 
parameter(s) of $f(\phi)$:
\begin{equation}
\kappa^2_Df(\phi_b)={{4(n-1)}\over\Delta}r^{-1}_bh^{{2\over{n\Delta}}
+{{n-3}\over{n-1}}}_-(r_b)h^{1\over 2}_+(r_b),
\label{rel1}
\end{equation}
\begin{equation}
\kappa^2_Df^{\prime}(\phi_b)=-{{4(n-1)a}\over\Delta}r^{-1}_b
h^{{2\over{n\Delta}}+{{n-3}\over{n-1}}}_-(r_b)h^{1\over 2}_+(r_b),
\label{drvrel}
\end{equation}
respectively, where $\phi_b\equiv\phi(r_b)$.  We note that the following 
choice of $f(\phi)$ trivially satisfies these constraints:
\begin{equation}
f(\phi)=\sigma e^{-a\phi},
\label{brntens}
\end{equation}
where $\sigma$ is the $n$-brane tension.
The consistency of the boundary conditions resulting from Eqs. 
(\ref{eqn2},\ref{eqn3}) fixes the stabilized location of the 
$n$-brane to be
\begin{equation}
r^{n-1}_b={{4-\Delta}\over{4-2\Delta}}m\cosh^2\alpha.
\label{brnlctn}
\end{equation}
From this, we see that a real-valued brane position $0\leq r_b<r_+$ is 
possible when $0<\Delta<2$ or $\Delta\geq 4$ with $m>0$.  This result is in 
contrast to the case of the uncharged topological AdS black hole, in which 
the static brane configuration with $k=0$ is possible only when the black 
hole mass is zero \cite{br}.  We briefly discuss the limiting cases.  When 
$\Delta=4$, the stable position of the $n$-brane is at $r_b=0$, namely at the 
singularity.  When $\Delta=0$, the stable position of the $n$-brane is at 
the horizon $r^{n-1}_b=m\cosh^2\alpha$, meaning that the range of the 
transverse coordinate $\bar{z}$ is zero, i.e., this case is degenerate.  
When $\Delta=2$ the static configuration is not possible, since $r_b=\infty$ 
for such case.  For the bulk background of the uncharged 0-brane 
(\ref{uchrgdsol}) with $k=0$, the consistency of the boundary conditions 
resulting from Eqs. (\ref{eqn2},\ref{eqn3}) requires that $\Delta=2(n-1)$, 
meaning that as long as $\Delta=2(n-1)$ the static brane configuration for 
any values of the $n$-brane location $r_b$ is possible.  So, the static 
brane configuration in the background of the self-tuning solution, satisfying  
(\ref{poincrcndtn}), is possible only when $n=3$.  
The following fine-tuned value of the $n$-brane tension is obtained by 
substituting the expression for $r_b$ in Eq. (\ref{brnlctn}) into Eq. 
(\ref{rel1}) along with Eq. (\ref{brntens}):
\begin{equation}
\sigma=\left.{1\over\kappa^2_D}{{4(n-1)}\over\Delta}r^{-1}h^{-{{2(\Delta-n+1)}
\over{(n-1)\Delta}}}_-h^{1\over 2}_+\right|_{r=r_b},
\label{brntns}
\end{equation}
where $h_-(r_b)={{4-2\Delta}\over{4-\Delta}}\tanh^2\alpha$ and $h_+(r_b)=
{\Delta\over{\Delta-4}}$.  From this explicit expression for the brane 
tension, we see that real-valued nonzero brane tension is possible when 
$\Delta>4$ or $\Delta<0$.  So, in order to have the real-valued $n$-brane 
location and tension, the dilaton coupling parameter $a$ should be such that 
$\Delta>4$.  When $\Delta=4$, for which the $n$-brane is located at the 
singularity $r=0$, the $n$-brane tension becomes infinite.  When $\Delta=0$, 
for which the $n$-brane is located at the event horizon $r=r_+$, the 
$n$-brane tension is also infinite.

Finally, we study a massless bulk scalar field $\Phi$ satisfying the 
following equation of motion:
\begin{equation}
{1\over\sqrt{-G}}\partial_M(\sqrt{-G}G^{MN}\partial_N\Phi)=0.
\label{scleq}
\end{equation}
We take the following ansatz for the solution:
\begin{equation}
\Phi(x^{\mu},z)=u_m(x^{\mu})f_m(z),\ \ \ \ \ 
-A^{-2}\partial^2_tu_m+B^{-2}\delta^{ij}\partial_i\partial_ju_m=-m^2u_m.
\label{sclans}
\end{equation}
Then, Eq. (\ref{scleq}) reduces to the Sturm-Liouville equation:
\begin{equation}
\partial_z[AB^n\partial_z]f_m=m^2AB^nf_m.
\label{scleq2}
\end{equation}
In terms of the new $z$-dependent function $\tilde{f}_m=A^{1\over 2}B^{n\over 
2}f_m$, this equation takes the following zero energy eigenvalue Schr\"odinger 
equation:
\begin{equation}
-{{d^2\tilde{f}_m}\over{dz^2}}+V(z)\tilde{f}_m=0,
\label{scheq}
\end{equation}
with the potential
\begin{equation}
V(z)={1\over 2}{A^{\prime\prime}\over A}+{n\over 2}{B^{\prime\prime}\over B}
+{n\over 2}{A^{\prime}\over A}{B^{\prime}\over B}-{1\over 4}\left({A^{\prime}
\over A}\right)^2+{{n(n-2)}\over 4}\left({B^{\prime}\over B}\right)^2+m^2.
\label{pot}
\end{equation}
This equation can be rewritten in the following convenient form:
\begin{equation}
\left[\bar{Q}Q+m^2\right]\tilde{f}_m=0,
\label{sqmeq}
\end{equation}
where
\begin{equation}
Q=-{d\over{dz}}+{{d\ln A^{1\over 2}B^{n\over 2}}\over{dz}},\ \ \ \ \ 
\bar{Q}={d\over{dz}}+{{d\ln A^{1\over 2}B^{n\over 2}}\over{dz}}.
\label{defq}
\end{equation}
So, the zero mode ($m=0$) annihilated by $Q$ is
\begin{equation}
\tilde{f}_0=A^{1\over 2}B^{n\over 2},
\label{kkzmd}
\end{equation}
from which we see that the Kaluza-Klein zero mode $\Phi(x^{\mu},z)=u_0
(x^{\mu})f_0(z)=u_0(x^{\mu})A^{-{1\over 2}}B^{-{n\over 2}}\tilde{f}_0$ of 
the bulk scalar field is independent of $z$, just as in the symmetrically 
warped brane world case.  It is straightforward to show that such result 
continues to hold for other bosonic fields.   In order for the eigenvalue 
problem for the Sturm-Liouville equation to be well-defined, appropriate 
boundary conditions have to be satisfied.  For the Sturm-Liouville equation 
of the form (\ref{scleq2}), the eigenvalue $m^2$ is real and the 
eigenfunctions $f_m$ with different eigenvalues are orthogonal w.r.t. the 
weighting function $w(z)=AB^n$, provided the operator ${\cal L}=\partial_z
[AB^n\partial_z]$ is self-adjoint.  The operator ${\cal L}$ is self-adjoint, 
in general when the boundary condition $\left.AB^n(f^{\prime}_mf_{m^{\prime}}-
f^{\prime}_{m^{\prime}}f_m)\right|^{z_H}_{z_b}=0$ is satisfied.  There are 
many ways in which this boundary condition can be satisfied.  One of possible 
ways would be to first write the eigenfunction as $f_n(z)=F_m(m^2z)$ and 
require that $F_m(m^2z)$ vanishes at the boundaries.  This requirement 
determines the eigenvalues $m^2$ satisfying the required boundary condition. 

\end{sloppypar}

\end{document}